\documentclass[prl,reprint,twocolumn,10pt,showpacs,superscriptaddress,nofootinbib]{revtex4-1}
\usepackage{amsmath}
\usepackage{amsthm}
\usepackage{graphicx}
\usepackage{times}
\usepackage{amssymb}
\usepackage{hyperref}
\hypersetup{
colorlinks=true,
linkcolor=red,
citecolor=red,}
\usepackage{textcomp}
\usepackage{xcolor}
\usepackage{enumerate}
\usepackage{multirow}
\usepackage{tabularx}
\usepackage{dcolumn}
\usepackage[mathscr]{euscript}
\usepackage{mathtools}
\usepackage{needspace}
\usepackage[mathscr]{euscript}

\begin{document}
\title{Ultrafine Entanglement Witnessing}
\author{Farid Shahandeh}
\email{Electronic address: f.shahandeh@uq.edu.au}
 \affiliation{Centre for Quantum Computation and Communication Technology, School of Mathematics and Physics, University of Queensland, St Lucia, Queensland 4072, Australia}
\author{Martin Ringbauer}
\email{Electronic address: m.ringbauer@uq.edu.au}
 \affiliation{Centre for Quantum Computation and Communication Technology, School of Mathematics and Physics, University of Queensland, St Lucia, Queensland 4072, Australia}
\affiliation{Centre for Engineered Quantum Systems, School of Mathematics and Physics, University of Queensland, St Lucia, Queensland 4072, Australia}
\author{Juan C. Loredo}
 \affiliation{Centre for Quantum Computation and Communication Technology, School of Mathematics and Physics, University of Queensland, St Lucia, Queensland 4072, Australia}
\affiliation{Centre for Engineered Quantum Systems, School of Mathematics and Physics, University of Queensland, St Lucia, Queensland 4072, Australia}
\author{Timothy C. Ralph}
 \affiliation{Centre for Quantum Computation and Communication Technology, School of Mathematics and Physics, University of Queensland, St Lucia, Queensland 4072, Australia}
\date{\today}

\begin{abstract}
Entanglement witnesses are invaluable for efficient quantum entanglement certification without the need for expensive quantum state tomography. Yet, standard entanglement witnessing requires multiple measurements and its bounds can be elusive as a result of experimental imperfections.
Here we introduce and demonstrate a novel procedure for entanglement detection which seamlessly and easily improves any standard witnessing procedure by using additional available information to tighten the witnessing bounds.
Moreover, by relaxing the requirements on the witness operators, our method removes the general need for the difficult task of witness decomposition into local observables.
We experimentally demonstrate entanglement detection with our approach using a separable test operator and a simple fixed measurement-device for each agent.
Finally we show that the method can be generalized to higher-dimensional and multipartite cases with a complexity that scales linearly with the number of parties.
\end{abstract}

%\pacs{42.50.-p, 03.67.-a, 03.65.Ta, 42.50.Ex, 03.65.Ud}

\maketitle

%======================================================
%======================================================
%				Introduction

%\paragraph*{Introduction.---}
Quantum entanglement provides many advantages beyond classical limits, including quantum communication, computation, and information processing~\cite{Nielsen,Horodecki09}. 
Yet, determining whether a given quantum state is entangled or not is a theoretically and experimentally challenging task~\cite{Gurvits03,Gharibian10}. In particular, the ideal approach of reconstructing the full quantum state via quantum tomography is practically infeasible for all but the smallest systems.

\begin{figure}[h]
  \includegraphics[width=0.8\columnwidth]{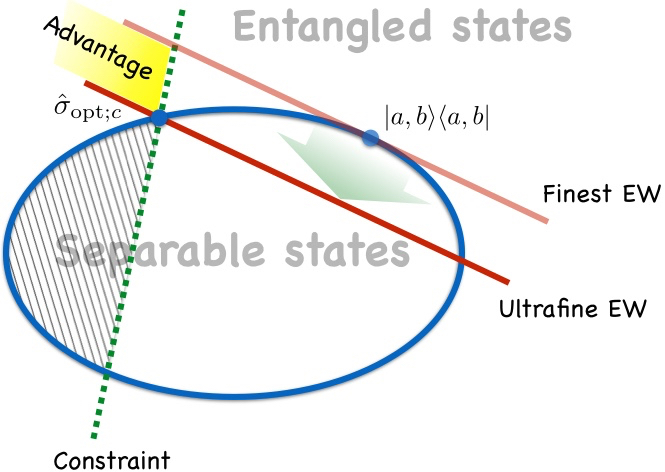}
  \vspace{-0em}
  \caption{(Color online) Conceptual idea of ultrafine entanglement witnessing. In a standard witnessing,  the finest entanglement witness (EW) is obtained by shifting a {\it test} operator so that its corresponding hyperplane becomes tangent to the set of separable states, and thus, optimal with respect to this set. This is termed the \emph{finest EW}. However, additional information or constraints on the quantum states under investigation can effectively reduce the size of the set of candidate separable states (the hashed subset). Our technique takes this into account to provide an \emph{ultrafine EW} that is tangent to this reduced set of separable states. This, in general, leads to an advantage over the standard procedure (the yellow region). By varying the constraint one can scan a large range of entangled states.}
  \label{fig:WPC}
\end{figure}

An elegant solution to this problem, known as entanglement witnessing, relies on the geometry of the set of all non-entangled (separable) quantum states~\cite{Horodecki09,Horodecki96,Horodecki01,Chruscinski14}.
Since these states form a convex set, it is always possible to find a hyperplane such that a given entangled state lies on one side of the hyperplane, while all separable states are on the other side, see Fig.~\ref{fig:WPC}.
This hyperplane is a so-called \emph{entanglement witness} (EW) and corresponds to a joint observable that has a bounded expectation value over all separable quantum states.
Any quantum state that produces a value beyond the bound must therefore be entangled.
This simplification, however, comes at a cost: 
first, different entangled states in general require different entanglement witnesses to be detected; 
second, not every EW can be practically realized, i.e.,\ can be decomposed in terms of operators corresponding to available local measurement devices (See also Refs.~\cite{Chruscinski14,Yu05,Gholipour16} for examples of the reverse procedure: constructing EWs from local observables);
third, when such a decomposition is possible, it might require multiple measurement devices (with multiple settings) to be implemented;
and fourth, witnessing bounds can be elusive in the presence of experimental imperfections.
Consequently, one would ideally like to construct EWs that have a simple decomposition and, at the same time, detect a large set of entangled states.

There are three main techniques to improve EWs.
First, to use nonlinear EWs by addition of nonlinear terms to the original witness operator~\cite{Guhne06}.
Second, by using collective measurement of EWs on multiple copies of quantum states~\cite{Horodecki03}.
Third, by optimizing a given witness to tighten the bound on the statistics of separable states as much as possible~\cite{Lewen00,Sperling09}.
The latter, which we refer to as standard entanglement witnessing (SEW), is the most common procedure and enjoys the property that it can be used as a complementary procedure to the first two techniques.
In SEW one first evaluates the supremum (infimum) expectation value of the witness observable for all separable states. The witness operator is then decomposed into local measurements, such that the expectation value can be computed from the combined local measurement statistics. 
A comparison against the corresponding upper (lower) bound for separable states establishes the entanglement of the tested state. Crucially, a significant amount of information from these measurements remains unused by combining the statistics.

In this Letter, we introduce and demonstrate a novel approach for witnessing quantum entanglement that makes use of this otherwise unused additional information to seamlessly and easily enhance any existing witnessing protocol.
Our method, which we call \emph{ultrafine entanglement witnessing} (UEW), relaxes the requirements on the test operators, which facilitates entanglement detection of a wide range of states, even if no witness decomposition in the common sense is provided.
Moreover, our approach makes it possible to detect entanglement using only a simple measurement device for each party with the minimal number of measurements.
We implement this technique experimentally on two-qubit entangled states shared by Alice and Bob, each of whom has access to a fixed three-outcome measurement device. 
Finally, we show that UEW can straightforwardly be extended to multipartite scenarios, with an experimental complexity that scales linearly with the number of parties involved.

%======================================================
%======================================================
%			Standard Witnessing Procedure

%\paragraph*{Standard Entanglement Witnessing.---}

Standard entanglement witnessing relies on the fact that the set of all separable states, $\mathcal{S}_{\rm sep}$, is the collection of all convex combinations of pure product states. 
As a fruitful consequence of this convexity, we can identify quantum states outside $\mathcal{S}_{\rm sep}$ (i.e.,\ entangled states) as follows. 
The Hahn-Banach theorem implies that for every entangled state $\hat{\varrho}$ there exists a hyperplane cutting through the state space that separates $\hat{\varrho}$ from the set of separable states, see Fig.~\ref{fig:WPC}. 
Mathematically, there exists a Hermitian operator $\hat{W}$ such that ${\rm Tr}\hat{\sigma}\hat{W}\geq 0$ for all $\hat{\sigma}\in\mathcal{S}_{\rm sep}$, while ${\rm Tr}\hat{\varrho}\hat{W}< 0$~\cite{Horodecki96}. 
The operator $\hat{W}$ is called an \emph{entanglement witness}.

Powerful EWs are most commonly constructed by optimizing a Hermitian (and possibly completely positive) \emph{test} operator $\hat{L}$ over the set of separable states as~\cite{Lewen00,Sperling09}
\begin{equation}
\label{switconst}
\hat{W}:=g_{\rm s}\hat{I}-\hat{L} ,
\end{equation}
where $\hat{I}$ is the identity operator, and $g_{\rm s}=\sup\{{\rm Tr}\hat{L}\hat{\sigma}:\hat{\sigma}\in\mathcal{S}_{\rm sep}\}$. 
Indeed, it is sufficient to optimize only over pure product states $|a,b\rangle$~\cite{Sperling09}.
One can also employ a similar recipe using the infimum value $g_{\rm i}=\inf\{{\rm Tr}\hat{L}\hat{\sigma}:\hat{\sigma}\in\mathcal{S}_{\rm sep}\}$.
This optimization procedure can be geometrically understood as translating the hyperplane corresponding to the test operator until it is tangent to the set of separable states, as illustrated in Fig.~\ref{fig:WPC}. Hence, there exists a pure product state, called the \emph{optimal point}, for which $\langle a,b|\hat{W}|a,b\rangle = 0$~\cite{Chruscinski14,Shultz16}. 
The resulting EW is said to be the \emph{finest} witness in the sense that any further shift of its corresponding hyperplane will lead to an operator whose expectation value becomes negative for some separable states, thus, violating the proper witnessing conditions~\cite{Lewen00,Sperling09}.

%======================================================
%======================================================
%			Constrained Witnessing Procedure

%\paragraph*{Ultrafine Entanglement Witnessing.---}
It is, however, possible to significantly increase the detection power of any test operator by taking into account additional constraints and information about the states under investigation, which effectively reduces the size of the set of viable separable states, see Fig.~\ref{fig:WPC}.
These constraints reflect physical restrictions on the measurement statistics that can be produced by separable states in certain situations. Similar considerations have previously been applied in the context of non-Gaussianity detection~\cite{Filip11}.
Consider, for example, a system composed of two spin-$1/2$ particles, which, in a measurement along the $z$-axis, are always found either both with spin up, or both with spin down. 
There is a large number of separable states that cannot produce such statistics and can thus be excluded from the optimization procedure for any witness aiming to detect the potential entanglement.
Crucially, the required information about the state is already available, but not used, in almost every standard witnessing experiment. 

Consider a Hermitian operator $\hat{C}\neq\hat{W}$ corresponding to some physical observable. 
Just like a witness, $\hat{C}$ corresponds to a hyperplane cutting through the space of all quantum states, splitting it into two half-spaces $\mathcal{S}_{c}:=\{\hat{\varrho}:{\rm Tr}\hat{C}\hat{\varrho}\leq c\}$ and $\mathcal{S}_{\tilde{c}}:=\{\hat{\varrho}:{\rm Tr}\hat{C}\hat{\varrho}\geq c\}$, where $c$ is a real-valued free parameter.
Depending on the choice of constant $c$, the hyperplane $\hat{C}$ may or may not cut through the set of separable states, defining the two closed convex subsets $\mathcal{S}_{{\rm sep};c}:=\mathcal{S}_{\rm sep}\cap\mathcal{S}_{c}=\{\hat{\sigma}:\hat{\sigma}\in \mathcal{S}_{\rm sep} \text{ and } {\rm Tr}\hat{C}\hat{\sigma}\leq c\}$ and $\mathcal{S}_{{\rm sep};\tilde{c}}:=\mathcal{S}_{\rm sep}\cap\mathcal{S}_{\tilde{c}}=\{\hat{\sigma}:\hat{\sigma}\in \mathcal{S}_{\rm sep} \text{ and } {\rm Tr}\hat{C}\hat{\sigma}\geq c\}$.
Clearly, whenever one of the sets is empty the other one coincides with the set of all separable states, and hence, our method reduces to SEW.
Therefore, in the following, we will consider parameter values for which both $\mathcal{S}_{{\rm sep};c}$ and $\mathcal{S}_{{\rm sep};\tilde{c}}$ are nonempty.
Using the test operator $\hat{L}$ one can now construct two EWs, $\hat{W}_c$ and $\hat{W}_{\tilde{c}}$, optimal to the sets $\mathcal{S}_{{\rm sep};c}$ and $\mathcal{S}_{{\rm sep};\tilde{c}}$, respectively, by replacing $g_s$ of Eq.~\eqref{switconst} with $g_{c}=\sup\{{\rm Tr}\hat{L}\hat{\sigma}:\hat{\sigma}\in\mathcal{S}_{{\rm sep};c}\}$ and $g_{\tilde{c}}=\sup\{{\rm Tr}\hat{L}\hat{\sigma}:\hat{\sigma}\in\mathcal{S}_{{\rm sep};\tilde{c}}\}$.
Consequently, a state $\hat{\varrho}$ is entangled if

\begin{equation}
\label{DetConds}
\begin{split}
{\rm Tr}\hat{C}\hat{\varrho}\leq c \wedge {\rm Tr}\hat{W}_{c}\hat{\varrho}<0, \text{~or,~} {\rm Tr}\hat{C}\hat{\varrho}\geq c \wedge {\rm Tr}\hat{W}_{\tilde{c}}\hat{\varrho}<0 .
\end{split}
\end{equation}

\paragraph*{Lemma~1.}
Given a test operator $\hat{L}$ with optimal points to the sets $\mathcal{S}_{\rm sep}$ and $\mathcal{S}_{{\rm sep};X}$ as $|a,b\rangle$ and $\hat{\sigma}_{\rm opt;X}$ for $X=c,\tilde{c}$, respectively,
\begin{enumerate}[(i)]
\item If $\langle a,b|\hat{C}|a,b\rangle \leq c$, then $g_c = g_{\rm s}$ and $\hat{\sigma}_{{\rm opt};c} = |a,b\rangle\langle a,b|$, i.e.,\ $\hat{W}_{c} = \hat{W}$. 
Furthermore, ${\rm Tr}\hat{C}\hat{\sigma}_{{\rm opt};\tilde{c}}=c$.
\item If $\langle a,b|\hat{C}|a,b\rangle \geq c$, then $g_{\tilde{c}} = g_{\rm s}$ and $\hat{\sigma}_{{\rm opt};\tilde{c}}{=}|a,b\rangle\langle a,b|$, i.e.,\ $\hat{W}_{\tilde{c}}=\hat{W}$.
Furthermore, ${\rm Tr}\hat{C}\hat{\sigma}_{{\rm opt};c}=c$.
\end{enumerate}
We point the interested reader to the Supplemental Material~\cite{Supp} for the proof of Lemma 1. 
Lemma~1 shows that the optimal point from SEW remains optimal for one of the two sets $\mathcal{S}_{{\rm sep};X}$, while for the other set the optimal point lies on the hyperplane $\hat{C}$, as visualized in Fig.~\ref{fig:WPC}. 
This in particular implies that for a given $c$ one of the conditions in Eq.~\eqref{DetConds} is advantageous over SEW.
In addition, Eq.~\eqref{DetConds} together with Lemma~1 imply that UEW and SEW are equivalent only in the special case that the constraint value $c$ is chosen exactly to match the expectation value of the constraint operator in the SEW optimal point, i.e., for $c=\langle a,b|\hat{C}|a,b\rangle$.
This, in turn, implies that our strategy never performs worse than SEW.
Accordingly, we also obtain the following useful results the proofs of which are provided in the Supplemental Material~\cite{Supp}.

\paragraph*{Theorem~1.}
For a given constraint value $c$, the optimal state $\hat{\sigma}_{{\rm opt};X}\in\mathcal{S}_{{\rm sep};X}$ to the test operator $\hat{L}$ is a pure state with ${\rm Tr}\hat{C}\hat{\sigma}_{{\rm opt};X}=c$.

\paragraph*{Theorem~2.}
The necessary condition for the separable operators $\hat{C}$ and $\hat{L}$ to detect entanglement via UEW is that $[\hat{C},\hat{L}]\neq0$.

\paragraph*{Corollary~1.} 
If $\hat{C}=\hat{C}^{\rm A}\otimes\hat{C}^{\rm B}$ and $\hat{L}=\hat{L}^{\rm A}\otimes\hat{L}^{\rm B}$ are product operators, then $\hat{C}^{\rm Y}$ and $\hat{L}^{\rm Y}$ ($\rm Y=A$,$\rm B$) must be jointly non-measurable operators.

In SEW it is necessary that the test operator $\hat{L}$ has an entangled eigenspace, since otherwise the supremum (and infimum) expectation values could be obtained by separable eigenstates.
Theorem~2 and Corollary~1 show that our approach relaxes this requirement on the test operators and can be implemented with two separable (or even product) Hermitian operators. 
Notably, Corollary~1 implies that each party must use a measurement device with at least three outcomes, independent of the Hilbert space dimension of the system.
This property makes UEW very efficient for experimental applications by significantly reducing the number of measurements required for detection of particular classes of bipartite and multipartite entangled states.

%======================================================
%======================================================
%		       	UFW in Practice

%\paragraph*{UEW in Practice.---}
Consider two parties, Alice and Bob, each of whom has access to a single measurement device $\mathcal{M}^{\rm A}=\{\hat{\Pi}^{\rm A}_{i}\}_{i=1}^{n}$, and $\mathcal{M}^{\rm B}=\{\hat{\Pi}^{\rm B}_{i}\}_{i=1}^{m}$, respectively. In the simplest scenario, Alice and Bob can use their measurement devices to detect a range of entangled states using the following protocol.

\begin{figure}[h]
  \includegraphics[width=\columnwidth]{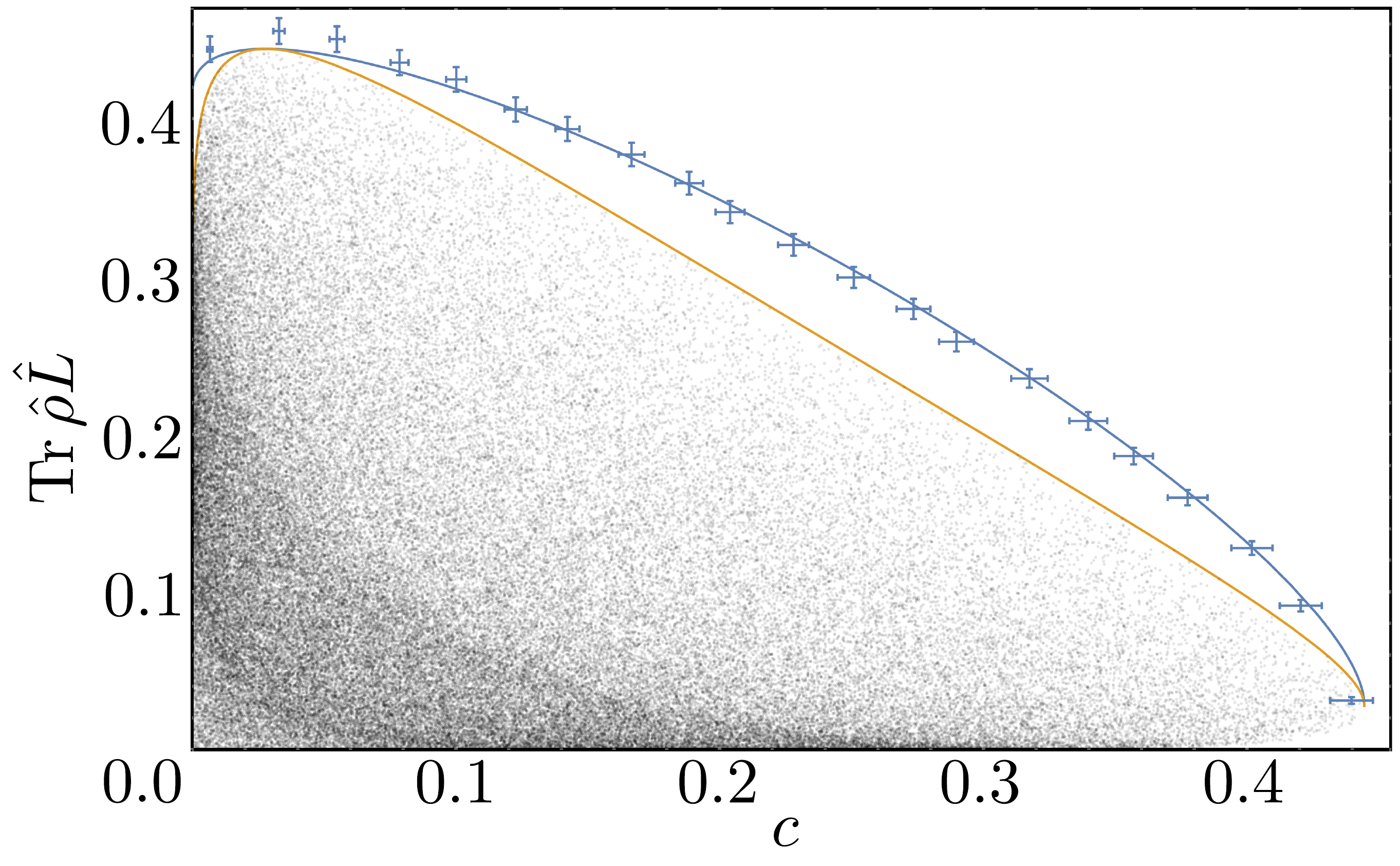}
  \vspace{-2em}
  \caption{(Color online) Experimental results for ultrafine entanglement witnessing using the single measurement device defined in Eqs.~\eqref{POVMs} and~\eqref{ExOps} with $x=2/3$ and $\theta=0$.
  The separability curve, which represents the largest expectation values of the test operator obtainable from separable states, is shown in orange, while the maximal values obtainable with entangled states are represented by the blue curve. 
  The blue data points correspond to 21 equispaced entangled states and include $3\sigma$ error bars.
The black dots, obtained from randomly sampled pure product states with uniform distributions over the local Bloch spheres, illustrate the density of separable states with respect to the test and constraint operators. Note that the point where the orange and blue curves meet is the optimal point one would obtain for $\hat{L}$ in SEW. This highlights that SEW is unable to detect any entanglement with our choice of a separable test operator.}
  \label{fig:results}
\end{figure}

\begin{enumerate}[(i)]
\item Choose a constraint operator of the form $\hat{C}=\hat{\Pi}^{\rm A}_{i}\otimes\hat{\Pi}^{\rm B}_{i}$.
\item Choose a test operator of the form $\hat{L}=\hat{\Pi}^{\rm A}_{j}\otimes\hat{\Pi}^{\rm B}_{j}$ for $j\neq i$; note that the conditions of Theorem~2 and Corollary~1 must apply to $\hat{C}$ and $\hat{L}$.
\item For each value $c$, compute $g(c)=\sup\{\langle a,b|\hat{L}|a,b\rangle:|a,b\rangle\in\mathcal{S}_{\rm sep}\}$ constrained to $\langle a,b|\hat{C}|a,b\rangle=c$.
\item The result is the concave curve of $g(c)$ versus $c$, called the {\it separability curve}. 
Any point corresponding to a quantum state above this curve indicates either of the conditions in Eq.~\eqref{DetConds}, and thus implies its entanglement.
\end{enumerate}
Corollary~1 implies that $\hat{\Pi}^{\rm Y}_{i}$ and $\hat{\Pi}^{\rm Y}_{j}$ must not be jointly measurable and thus, $\mathcal{M}^{\rm A}$ and $\mathcal{M}^{\rm B}$ cannot be dichotomic measurements.
In the case where Alice and Bob use the same three-outcome measurement device with $\hat{C}$ defined as above, there are six different separability curves. 
One of these curves is shown in Fig.~\ref{fig:results} together with the expectation and experimental observations for a family of entangled states. We emphasize here that, in general, arbitrary POVM elements can be combined to form complex constraints and test operators. Moreover, one might consider using multiple constraints, which would lead to \emph{separability hypersurfaces}. 
Hence, there is a large number of different possible ways to implement UEW with a single measurement device.

\begin{figure}[h]
  \includegraphics[width=0.95\columnwidth]{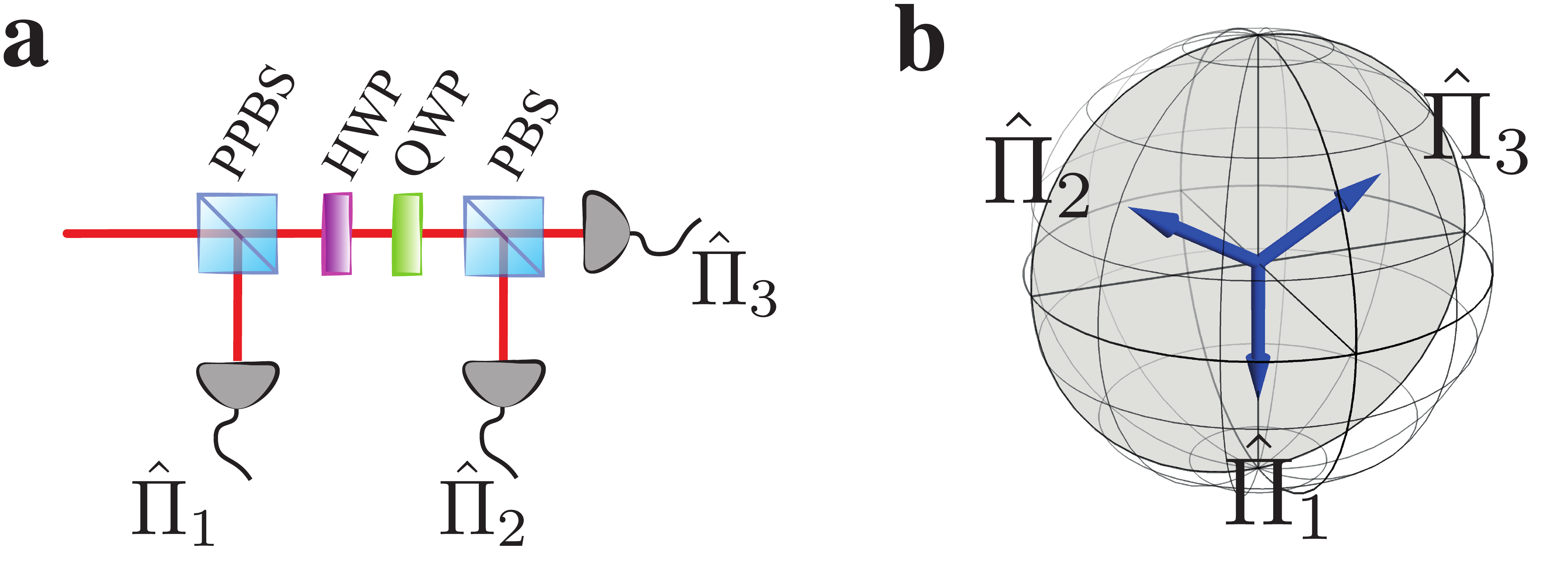}
  \vspace{-1em}
  \caption{(Color online) (a) Experimental implementation of the three-outcome qubit measurement of Eq.~\eqref{POVMs}.
  The first POVM element $\hat\Pi_1$ is implemented directly by a partially polarizing beam splitter (PPBS) with reflection coefficients $r_H=0$ for horizontal and $r_V=2/3$ for vertical polarization. 
  The other POVM elements $\hat\Pi_2$ and $\hat\Pi_3$ are implemented using a set of quarter-waveplate (QWP), half-waveplate (HWP) and a polarizing beam splitter (PBS).
  (b) Visualization of our three-outcome POVM in the \textsc{xz}-plane of the Bloch sphere.}
\label{fig:setup}
\end{figure}

Experimentally we consider two-qubit states encoded in the polarization of single photons, shared between Alice and Bob. They are both equipped with a three-outcome measurement device as shown in Fig.~\ref{fig:setup}, which implements the POVM elements
\begin{equation}
\label{POVMs}
\hat{\Pi}_1=x|V\rangle\langle V|,~
\hat{\Pi}_2=|\chi^+\rangle\langle \chi^+|,~
\hat{\Pi}_3=|\chi^-\rangle\langle \chi^-|,
\end{equation}
where $|\chi^\pm\rangle = \frac{1}{\sqrt{2}}|H\rangle \pm e^{i\theta}\sqrt{\frac{1-x}{2}}|V\rangle$ with an arbitrary phase $\theta$ and $\sum_{i=1}^3 \hat{\Pi}_i=\hat{I}$. 
Alice and Bob then choose the test and constraint operators as 
\begin{equation}
\label{ExOps}
\hat{L}=\hat{\Pi}^{\rm A}_2\otimes\hat{\Pi}^{\rm B}_2,\quad \hat{C}=\hat{\Pi}^{\rm A}_1\otimes\hat{\Pi}^{\rm B}_1 ,
\end{equation}
which implies that $0\leq c\leq x^2$. 
The corresponding separability curve for the case $x=2/3$ is shown in Fig.~\ref{fig:results}, together with a density plot of $10^{5}$ separable states, randomly sampled from the uniform distribution of pure states on Alice's and Bob's local Bloch spheres. 
Equations~\eqref{POVMs} and~\eqref{ExOps} imply that in our experiment the constraint corresponds to a limit on the vertical polarization component.

Starting from a general pure state we find that the maximal violation of the bound is obtained by states of the form
\begin{equation} 
\label{OptEnt}
|\phi\rangle = \alpha |HH\rangle + \beta e^{-i\theta} |HV\rangle + \gamma e^{-i\theta} |VH\rangle + \delta |VV\rangle ,
\end{equation}
with $\alpha,\beta,\gamma,\delta\in \mathbb{R}$ satisfying $\alpha^2+\beta^2+\gamma^2+\delta^2=1$.
Equation~\eqref{POVMs} thus implies that $\delta = \sqrt{c}/x$ for a given value of $c$. 
Maximizing the expectation value of the test operator is then equivalent to maximizing the overlap $\langle\chi^+\chi^+|\phi\rangle=[\alpha + \sqrt{1-x}(\beta + \gamma) + \sqrt{c}((1-x)/x)]/2$. Since the last term is independent of the chosen state, we can assume $\beta=\gamma$ which reduces the problem to maximizing $\alpha/2 + \sqrt{1-x}\beta$ constrained to $\alpha^2+2\beta^2=1-c/x^2$. 
For $x=2/3$ one then obtains $\alpha_{\sup}=\sqrt{3(4-9 c)/20}$, $\beta_{\sup}=\sqrt{(4-9 c)/20}$, and the maximum expectation value of the test operator
\begin{equation}
\sup\{{\rm Tr}\hat{\varrho}_{\rm ent}\hat{L}:{\rm Tr}\hat{\varrho}_{\rm ent}\hat{C}=c\}=(\alpha_{\sup}+\sqrt{c}/2)^2.
\end{equation}
Note that these values are independent of $\theta$. Figure~\ref{fig:results} shows the theoretical maximal violation curve, together with our experimental results for $\theta=0$.

%======================================================
%======================================================
%			Improving SEWP

%\paragraph*{Improving SEW.---}

As we have seen, our approach allows for entanglement detection using separable test and constraint operators, which is not possible in SEW.
In addition, UEW can be used to seamlessly and easily improve any existing standard witnessing experiment:
Consider two parties implementing a standard EW using the test operator $\hat{L}$, decomposed into local POVM elements as $\hat{L}=\sum_i \beta_i \hat{\Pi}^{\rm A}_{i}\otimes\hat{\Pi}^{\rm B}_{i}$ with $\beta_i\in \mathbb{R}$. 
Their aim is to violate the inequality ${\rm Tr}\hat{L}\hat{\sigma}\leq g_{\rm s}$ using an entangled state $\hat{\varrho}$. 
In such a scenario UEW allows us to tighten the bound, thus making the entanglement detection more experimentally robust. 
This is achieved by choosing an arbitrary pair (or a subset) of POVM elements as a constraint, say $\hat{C}=\hat{\Pi}^{\rm A}_{1}\otimes\hat{\Pi}^{\rm B}_{1}$, and computing ${\rm Tr}\hat{C}\hat{\varrho}=c$ using already measured data.
Then, one re-optimizes $\hat{L}$ over the set of pure product states with $\langle a,b|\hat{C}|a,b\rangle = c$ to obtain the new tighter bound $g(c)\leq g_{\rm s}$ and tests the inequality ${\rm Tr}\hat{L}\hat{\sigma}\leq g(c)$. 
Notably, there are many (in fact, infinitely many) constraints that could be constructed from the initially measured POVM elements.

%======================================================
%======================================================
%			Multipartite Extension

%\paragraph*{Multipartite Extension.---}

We now show how the simple procedure for UEW outlined above can be directly extended to the multipartite scenario where a quantum state is shared between multiple parties.
SEW for this case has been demonstrated in theory and experiment in Refs.~\cite{Sperling13,Gehrke15}.
Consider a $N$-qubit system shared between $N$ agents, each of them having a three-outcome measurement device with POVM elements given by Eq.~\eqref{POVMs}.
Moreover, suppose that an arbitrary $k$-partitioning of the system has been chosen as $\mathbf{P}_k=(\mathcal{I}_1|\mathcal{I}_2|\cdots|\mathcal{I}_k)$, where each party $\mathcal{I}_i$ is a subset of the index set $\mathcal{I}=\{1,2,\dots,N\}$, containing $M_i$ agents (and hence, subsystems), so that $\sum_i {\rm card }\,\mathcal{I}_i=\sum_i M_i=N$.
Moreover, the list of parties is ordered such that $M_1\leq M_2\leq\cdots\leq M_k$.
Now, the agents chose the test and constraint operators as
\begin{equation}\label{MultiTestOp}
\hat{L} =\bigotimes_{i=1}^{k}\hat{L}_i=\bigotimes_{i\in\mathcal{I}_1}\hat{\Pi}^{(i)}_2\bigotimes_{i\in\mathcal{I}_2}\hat{\Pi}^{(i)}_2\cdots\bigotimes_{i\in\mathcal{I}_k}\hat{\Pi}^{(i)}_2,
\end{equation}
and 
\begin{equation}\label{MultiConstOp}
\hat{C} =\bigotimes_{i=1}^{k}\hat{C}_i=\bigotimes_{i\in\mathcal{I}_1}\hat{\Pi}^{(i)}_1\bigotimes_{i\in\mathcal{I}_2}\hat{\Pi}^{(i)}_1\cdots\bigotimes_{i\in\mathcal{I}_k}\hat{\Pi}^{(i)}_1,
\end{equation}
which implies that $0\leq c\leq x^N$.

As a proof-of-principles, suppose that $c=0$.
In the Supplemental Material~\cite{Supp}, we prove that the maximum separable bound for a partition $\mathbf{P}_k$ is given by
\begin{equation}\label{MaxMultiSep}
g (x;N,M_k) = (1-\frac{x}{2})^N - (1-\frac{x}{2})^{N-M_k}(\frac{1-x}{2})^{M_k}.
\end{equation}
Since the test and constraint operators are invariant under the exchange of agents between different parties, so is the bound $g(x;N,M_k)$. 
Three cases are of particular interest. 
First, if $M_k=N$, then no partitioning has been made and $g(x;N,N)$ represents the maximum expectation value of the test operator $\hat{L}$ over {\it all} $N$-partite quantum states and hence, this bound cannot be violated by any quantum state.
Second, if $M_k=N-1$ the resulting bound corresponds to the bipartitions for which there is one subsystem in one party and $N-1$ subsystems in the other. One can easily see that for any bipartition with $M_k<N-1$, $g (x;N,M_k)<g (x;N,N-1)$.
Consequently, any state violating this bound is entangled within all bipartitions and thus, genuinely $N$-partite entangled.
Finally, if $M_k=1$ each party constitutes one agent corresponding to the partition with the highest resolution, i.e., $\mathbf{P}_{N}$. Thus, any state violating the bound $g(x;N,1)$ is partially entangled.

\begin{figure}[h]
  \includegraphics[width=\columnwidth]{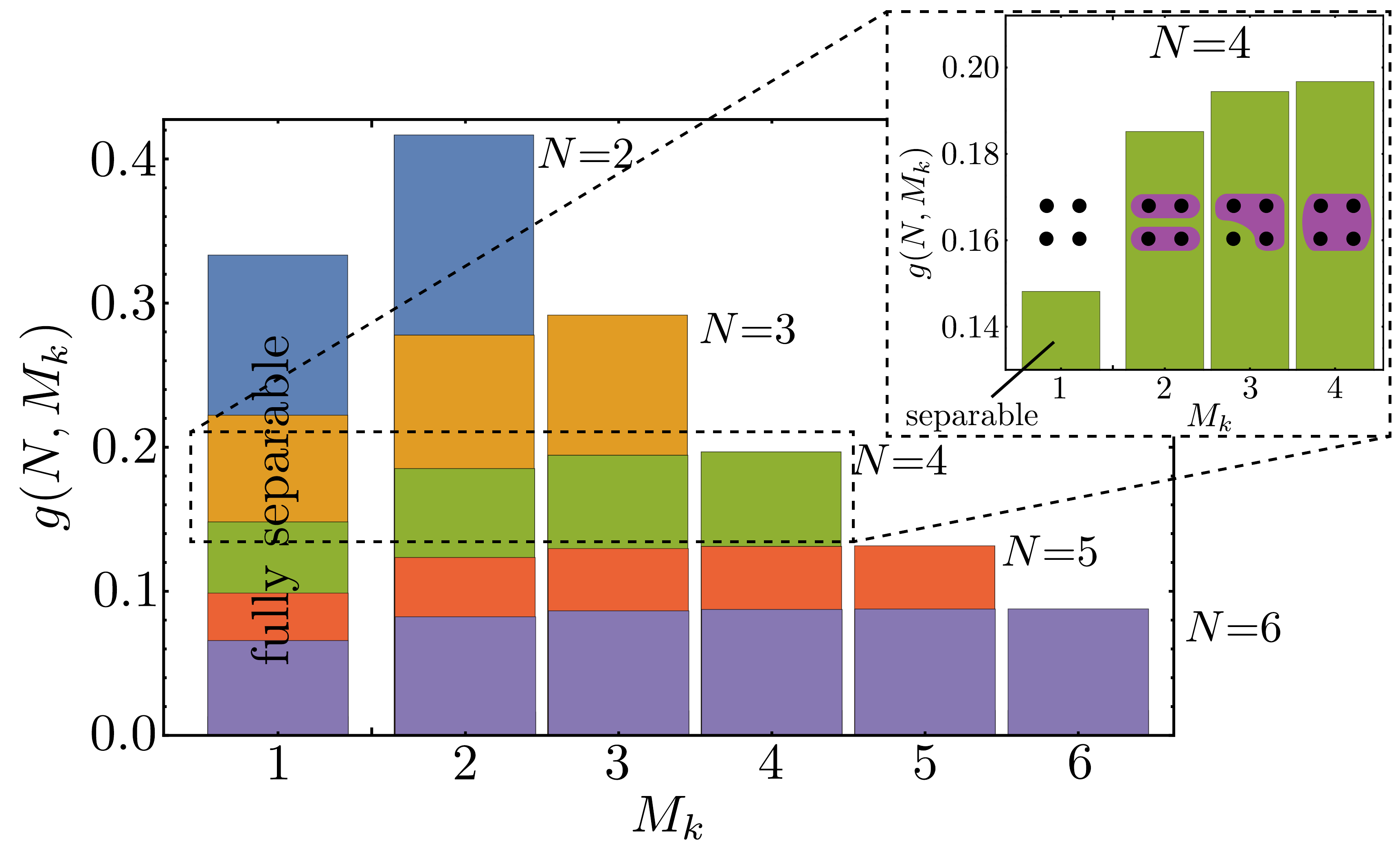}
  \vspace{-2em}
  \caption{(Color online) The bounds $g (N,M_k)$ versus the cardinality of the largest party $M_k$ for $2\leq N\leq 6$, where each $N$ is shown in a different color. 
For a state of interest with $c=0$, a violation of the bound $g (N,1)$ proves its (partial) entanglement, while a violation $g (N,N-1)$ proves its genuine $N$-partite entanglement. The inset shows the case $N=4$ with the four black dots representing the four subsystem and the purple shaded boxes visualize the maximally allowed entangled subset.}
  \label{fig:multipartite}
\end{figure}

Figure~\ref{fig:multipartite} shows $g (N,M_k):=g (\frac{2}{3};N,M_k)$ versus the cardinality of the largest party $M_k$ for $2\leq N\leq 6$. 
As $N$ increases it becomes increasingly harder to detect genuine $N$-partite entanglement with the simplest version of our approach. 
However, detecting partial entanglement by violating the bound $g (N,1)$ remains experimentally feasible for larger $N$. 
This example shows that our approach can be extended to the multipartite case, where it allows for simple entanglement detection with a number of measurements that scales as $3N$ with the number of agents $N$, as opposed to the tomographic methods or Bell tests for which the number of measurements scales exponentially with the number of qubits~\cite{Toth05}.
In fact, current EWs require at least $d+1$ measurements for each agent, where $d$ is the minimum Hilbert space dimension of the subsystems~\cite{Guhne03,Sperling13}, while our technique provides the possibility of detecting entangled states using only three-outcome measurements independent of the Hilbert space dimensionality.

%======================================================
%======================================================
%			Discussion and Conclusion

%\paragraph*{Discussion and Conclusion.---}
In conclusion, we have introduced a novel procedure for witnessing quantum entanglement using additional information that is typically already available in a standard witnessing experiment.
Our ultrafine entanglement witnessing relaxes the requirements on the test operators and allows for entanglement detection with a much smaller number of measurements compared to the standard entanglement witnessing.
In particular, our procedure makes it possible to witness entanglement using a single fixed measurement-device for each party.
This is a considerable experimental simplification which potentially allows for faster and more precise detection of entanglement compared to the existing protocols. 
We have demonstrated this in practice for a family of two-qubit entangled states using two fixed three-outcome POVMs.
We also showed that our method always performs at least as well as the standard procedure and seamlessly and easily improves it.
We have described a scalable experimental protocol that generalizes to higher dimensional and multipartite quantum systems, and showed that, in its simplest form, the number of measurements required for this protocol scales linearly with the number of agents.

%======================================================
%======================================================
%				Acknowledgements

\paragraph*{Acknowledgements.--}
The authors gratefully acknowledge A. G. White for helpful discussions and T. Vulpecula for experimental assistance. This project was supported by the ARC Centres for Quantum Computation and Communication Technology (CE110001027) and Engineered Quantum Systems (CE110001013).

\appendix
\begin{widetext}

\section*{Supplemental Material: \\
Ultrafine Entanglement Witnessing}

\subsection{Proof of Lemma~1}
We now prove part (i) of Lemma~1; part (ii) follows along the same line of proof.
Suppose that $\langle a,b|\hat{C}|a,b\rangle \leq c$.
Recalling that $\mathcal{S}_{{\rm sep};c}=\mathcal{S}_{\rm sep}\cap\mathcal{S}_{c}=\{\hat{\sigma}:\hat{\sigma}\in \mathcal{S}_{\rm sep} \text{ and } {\rm Tr}\hat{C}\hat{\sigma}\leq c\}$, the corresponding supremum expectation value of the test operator is $g_{\rm s}=\sup\{{\rm Tr}\hat{L}\hat{\sigma}:\hat{\sigma}\in\mathcal{S}_{\rm sep}\}=\sup\{{\rm Tr}\hat{L}\hat{\sigma}:\hat{\sigma}\in\mathcal{S}_{{\rm sep};c}\}=g_c$.

Now we prove that under the condition $\langle a,b|\hat{C}|a,b\rangle \leq c$ the optimal point $\hat{\sigma}_{{\rm opt};\tilde{c}}$ for the set $\mathcal{S}_{{\rm sep};\tilde{c}}$ must satisfy ${\rm Tr}\hat{C}\hat{\sigma}_{{\rm opt};\tilde{c}}=c$. 
For this, assume the contrary, that is, ${\rm Tr}\hat{C}\hat{\sigma}_{{\rm opt};\tilde{c}} \neq c$ and thus, ${\rm Tr}\hat{C}\hat{\sigma}_{{\rm opt};\tilde{c}} > c$.
Define the separable state $\hat{\chi}(p)=p\hat{\sigma}_{{\rm opt};\tilde{c}} + (1-p) |a,b\rangle\langle a,b|$ ($p\in[0,1]$). 
Since the two points $\hat{\sigma}_{{\rm opt};\tilde{c}}$ and $|a,b\rangle\langle a,b|$ are on two different sides of the hyperplane defined by $\hat{C}$, it is possible to find a $p^*$ such that $\hat{\chi}(p^*)$ lies on $\hat{C}$, i.e.,\ ${\rm Tr}\hat{C}\hat{\chi}(p^*) = c$ and $\hat{\chi}(p^*)\in\mathcal{S}_{{\rm sep};\tilde{c}}$.
Now, let us consider the position of $\hat{\chi}(p^*)$ with respect to $\hat{W}_{\tilde{c}}$.
We have
\begin{equation}\label{contradict}
\begin{split}
{\rm Tr}\hat{W}_{\tilde{c}}\hat{\chi}(p^*) &= p^* {\rm Tr}\hat{W}_{\tilde{c}}\hat{\sigma}_{{\rm opt};\tilde{c}} + (1-p^*) \langle a,b|\hat{W}_{\tilde{c}}|a,b\rangle\\
& = 0 + (1-p^*)( g_{\tilde{c}} - g_{\rm s}  ) < 0.
\end{split}
\end{equation}
The last inequality is a result of the convexity properties, as the supremum of a nonconstant convex (here, linear) function on a nonconstant convex set is strictly larger than its supremum over any subset of that convex set not equal to the set itself.
Equation~\eqref{contradict}, however, implies a contradiction to the witnessing property of $\hat{W}_{\tilde{c}}$ for the set $\mathcal{S}_{{\rm sep};\tilde{c}}$.
Thus, ${\rm Tr}\hat{C}\hat{\sigma}_{{\rm opt};\tilde{c}}= c$, as claimed.
\qed

\subsection{Proof of Theorem~1}
We now prove Theorem~1 in the case where condition (i) of Lemma~1 holds true. 
The same kind of proof can be used for the case where condition (ii) holds. 
First, without loss of generality, we restrict ourselves to the finite $d$-dimensional Hilbert space, as entanglement can always be detected in finite dimensions~\cite{Sperling2011finite}.
We also require the following Lemmas.
 
\paragraph*{Lemma~1~\cite{Cassinelli97}.---}
A given separable mixed state $\hat{\sigma}$ can be decomposed in such a way that it contains the pure product state $|a^*,b^*\rangle$ if $\hat{\sigma}^{-\frac{1}{2}}|a^*,b^*\rangle \neq |{\rm null}\rangle$.
\begin{proof}
Suppose that $\hat{\sigma}^{-\frac{1}{2}}|a^*,b^*\rangle = N^{-\frac{1}{2}}|x^*,y^*\rangle$ with $N=\langle a^*,b^*|\hat{\sigma}^{-1}|a^*,b^*\rangle$.
Choose an orthonormal basis set $\{|x_i,y_i\rangle\}_{i=1}^{d^2}$ such that for some $k$, $|x_k,y_k\rangle=|x^*,y^*\rangle$ and write 
\begin{equation}
\begin{split}
\hat{\sigma} & =\sum_i \hat{\sigma}^{\frac{1}{2}}|x_i,y_i\rangle\langle x_i,y_i|\hat{\sigma}^{\frac{1}{2}}\\
& = N |a^*,b^*\rangle\langle a^*,b^*| + \sum_{i\neq k} \hat{\sigma}^{\frac{1}{2}}|x_i,y_i\rangle\langle x_i,y_i|\hat{\sigma}^{\frac{1}{2}}.
\end{split}
\end{equation}
\end{proof}

\paragraph*{Lemma~2.---}
Any separable mixed quantum state $\hat{\sigma}\in\mathcal{S}_{\rm sep}$, which satisfies ${\rm Tr}\hat{C}\hat{\sigma}=c$ can be decomposed as $\hat{\sigma} = p|a,b\rangle\langle a,b| + (1-p)\hat{\sigma}_{\rm re}$ for some $p>0$ such that $|a,b\rangle\langle a,b|$ is a pure product state satisfying $\langle a,b|\hat{C}|a,b\rangle = c$.
The state $\hat{\sigma}_{\rm re}$ is the remainder.

\begin{proof}
Suppose that ${\rm Tr}\hat{C}\hat{\sigma}=c$ with $\hat{\sigma}\in\mathcal{S}_{\rm sep}$.

\begin{enumerate}[(i)]

\item If either $\mathcal{S}_{{\rm sep};c-\epsilon}=\emptyset$ or $\mathcal{S}_{{\rm sep};\widetilde{c+\epsilon}}=\emptyset$ for every $\epsilon>0$, then the corresponding hyperplane is tangent to the set of separable states.
As a result, $\hat{\sigma}$ belongs to a face of $\mathcal{S}_{\rm sep}$ and thus, it can be decomposed in terms of extremal points of that face, i.e., pure separable states satisfying $\langle a,b|\hat{C}|a,b\rangle = c$.

\item Assume that $\mathcal{S}_{{\rm sep};c-\epsilon}\neq\emptyset$ and $\mathcal{S}_{{\rm sep};\widetilde{c+\epsilon}}\neq\emptyset$ for some $\epsilon > 0$.
%Since $\hat{\sigma}$ has infinitely many decompositions in terms of separable states, 
Thus, there are infinitely many pure product states $|a,b\rangle$, such that $\langle a,b|\hat{C}|a,b\rangle = c_1 < c$, and infinitely many pure product states $|e,f\rangle$, such that $\langle e,f|\hat{C}|e,f\rangle = c_2 > c$.
Choose an arbitrary pair $|a,b\rangle$ and $|e,f\rangle$ from those states.
There necessarily exists a single parameter continuous family of unitary operators $\{\hat{U}(s)=(\hat{U}_{\rm A}\otimes\hat{U}_{\rm B})(s):s\in[0,1]\}$ such that $\hat{U}(0)=\hat{\mathbb{I}}$ and $\hat{U}(1)|a,b\rangle = |e,f\rangle$~\footnote{Such a family always exists: There exist (possibly many) Hamiltonians $\hat{H}_{\rm A}$ and $\hat{H}_{\rm B}$ such that $e^{i t_0(\hat{H}_{\rm A}+\hat{H}_{\rm B})}|a,b\rangle = |e,f\rangle$ for some characteristic time $t_0$.
Hence, one can define $s=t/t_0$ and $\hat{U}(s)=e^{i s(\hat{H}_{\rm A}+\hat{H}_{\rm A})}$}.
As a result of the continuity of $\hat{U}(s)$, the function $h(s)=\langle a,b|\hat{U}^\dag(s)\hat{C}\hat{U}(s)|a,b\rangle$ is continuous, $h(0)= c_1 < c$, and $h(1)=c_2 > c$~\footnote{Each $h(s)$ corresponds to a continuous path on the surface of the set of separable states connecting the points $|a,b\rangle\langle a,b|$ and $|e,f\rangle\langle e,f|$.}.
Therefore, there must exist a value of the parameter $s^*$ such that $h(s^*)=c$, and thus there exists a pure product state $|a^*,b^*\rangle = \hat{U}(s^*)|a,b\rangle$ for which $\langle a^*,b^*|\hat{C}|a^*,b^*\rangle = c$.
Now, since there are infinitely many $|a,b\rangle$'s, $|e,f\rangle$'s, and one parameter families $\{\hat{U}(s)\}$, there must exist infinitely many points $|a^*,b^*\rangle$.
Let us call the set of all such states $\mathcal{B}_{\rm sep;c}$.
Thus, one can always choose a complete set of vectors $\mathcal{O} = \{ |a_i,b_i\rangle\}_{i=1}^{d^2} $ from the set $\mathcal{B}_{\rm sep;c}$ such that $\langle a_i,b_i|\hat{C}|a_i,b_i\rangle = c$ for all $1\leq i \leq d^2$, because ${\rm card}\mathcal{O} < {\rm card}\mathcal{B}_{\rm sep;c}$.

The final step is to show that there must always exist at least one decomposition of $\hat{\sigma}$ which contains at least one $|a^*,b^*\rangle\langle a^*,b^*|$.
Equivalently, using Lemma~1 above, there must exist one $|a^*,b^*\rangle\langle a^*,b^*|$ for which $\hat{\sigma}^{-\frac{1}{2}}|a^*,b^*\rangle \neq |{\rm null}\rangle$.
To see this, assume that there is no  $|a^*,b^*\rangle$ such that $\hat{\sigma}^{-\frac{1}{2}}|a^*,b^*\rangle \neq |{\rm null}\rangle$, this means that $\hat{\sigma}^{-\frac{1}{2}}|a_i,b_i\rangle = 0$ for all elements of $\mathcal{O}$, and thus $\hat{\sigma}^{-\frac{1}{2}}=\hat{\sigma}^{-1} = \hat{0}$, i.e.,\ $\hat{\sigma}$ cannot be a quantum state, implying a contradiction.
\end{enumerate}

\end{proof}

We can now prove Theorem~1. Suppose that
\begin{equation}
\label{sigDec}
\hat{\sigma}_{{\rm opt};c}=\sum_i p_i |a_i,b_i\rangle\langle a_i,b_i|,
\end{equation}
where each $|a_i,b_i\rangle\langle a_i,b_i|$ is a pure product state and $\sum_i p_i = 1$.
Using Lemma~1, we know that ${\rm Tr}\hat{C}\hat{\sigma}_{{\rm opt};c}= c = \sum_i p_i c_i$ where $c_i =\langle a_i,b_i|\hat{C}|a_i,b_i\rangle$.
Thus, we can write
\begin{equation}\label{gcgc}
\begin{split}
g_c & = {\rm Tr} \hat{L}\hat{\sigma}_{{\rm opt};c}\\
& = \sum_i p_i \langle a_i,b_i|\hat{L}|a_i,b_i\rangle\\
& \leq \sum_i p_i \sup\{{\rm Tr}\hat{L}\hat{\sigma}:\hat{\sigma}\in\mathcal{S}_{{\rm sep};c_i}\}\\
& = \sum_i p_i {\rm Tr}\hat{L}\hat{\sigma}_{{\rm opt};c_i}, \qquad \hat{\sigma}_{{\rm opt};c_i}\in\mathcal{S}_{{\rm sep};c_i}\\
& =  {\rm Tr}\hat{L}\sum_i p_i\hat{\sigma}_{{\rm opt};c_i}\\
& \leq \sup\{{\rm Tr}\hat{L}\hat{\sigma}:\hat{\sigma}\in\mathcal{S}_{{\rm sep};c}\}\\
& = g_c.
\end{split}
\end{equation}
The first inequality follows from the fact that $|a_i,b_i\rangle\langle a_i,b_i|$ probably is not the optimal separable state over $\mathcal{S}_{{\rm sep};c_i}$.
The third equality is obtained by assuming that the optimal state of $\mathcal{S}_{{\rm sep};c_i}$ is $\hat{\sigma}_{{\rm opt};c_i}$.
The second inequality then follows by noticing that $\sum_i p_i\hat{\sigma}_{{\rm opt};c_i}\in\mathcal{S}_{{\rm sep};c}$.

Now, Eq.~\eqref{gcgc} implies that the two inequalities must be saturated.
Because of $\langle a_i,b_i|\hat{L}|a_i,b_i\rangle \leqslant \sup\{{\rm Tr}\hat{L}\hat{\sigma}:\hat{\sigma}\in\mathcal{S}_{{\rm sep};c_i}\}$, one necessarily obtains $g_{c_i}=\sup\{{\rm Tr}\hat{L}\hat{\sigma}:\hat{\sigma}\in\mathcal{S}_{{\rm sep};c_i}\}=\langle a_i,b_i|\hat{L}|a_i,b_i\rangle$ for each $i$, and, $g_c=\sup\{{\rm Tr}\hat{L}\hat{\sigma}:\hat{\sigma}\in\mathcal{S}_{{\rm sep};c}\}=\sum_i p_i \sup\{{\rm Tr}\hat{L}\hat{\sigma}:\hat{\sigma}\in\mathcal{S}_{{\rm sep};c_i}\}=\sum_i p_i g_{c_i}$.

Note that, we put no specific condition on the decomposition of $\hat{\sigma}_{{\rm opt};c}$ in Eq.~\eqref{sigDec}.
As a result, we use the fact that $\hat{\sigma}_{{\rm opt};c}$ is necessary inside the set $\mathcal{S}_{\rm sep}$ and we can use the Lemma above and say that we can always decompose $\hat{\sigma}_{{\rm opt};c}$ in the form $\hat{\sigma}_{{\rm opt};c} = p|a,b\rangle\langle a,b| + (1-p)\hat{\sigma}_{\rm re}$ for some $p>0$, $|a,b\rangle\langle a,b|$ being a pure product state satisfying $\langle a,b|\hat{C}|a,b\rangle = c$, and $\hat{\sigma}_{\rm re}$ being the remainder state.
Using the previous result, this implies $\hat{\sigma}_{{\rm opt};c}=|a,b\rangle\langle a,b|$, i.e.,\ the optimal point can always be chosen to be a pure state.
\qed

\subsection{Proof of Theorem~2}
The commutativity of $\hat{C}$ and $\hat{L}$ together with their separability imply that there exists a basis set of product operators $\{|i,j\rangle\}$ that diagonalizes both of them as $\hat{C}=\sum\mu_{ij}|i,j\rangle\langle i,j|$ and $\hat{L}=\sum\lambda_{ij}|i,j\rangle\langle i,j|$.
Suppose that there exists an entangled state $\hat{\varrho}$ such that ${\rm Tr}\hat{C}\hat{\varrho}=c$ and ${\rm Tr}\hat{L}\hat{\varrho}=l > g_{\rm s}$, with $g_{\rm s}$ being the supremum value attainable by separable states.
Consider the separable state
\begin{equation}
\hat{\sigma}=\sum \langle i,j|\hat{\varrho}|i,j\rangle |i,j\rangle\langle i,j|.
\end{equation}
It is clear that ${\rm Tr}\hat{C}\hat{\sigma}={\rm Tr}\hat{C}\hat{\varrho}=c$ and ${\rm Tr}\hat{L}\hat{\sigma}={\rm Tr}\hat{L}\hat{\varrho}=l > g$, implying a contradiction.
\qed

\subsection{Proof of Corollary~1}

Recall that, by Neumark's theorem, two operators $\hat{C}$ and $\hat{L}$ are jointly measurable if and only if they correspond to two operators $\hat{C}_0$ and $\hat{L}_0$ in an extended Hilbert space $\mathcal{H}_0$, respectively, satisfying $[\hat{C}_0,\hat{L}_0]=0$ such that $\hat{C}=\hat{P}\hat{C}_0\hat{P}$ and $\hat{L}=\hat{P}\hat{L}_0\hat{P}$ for an orthogonal projection operator $\hat{P}:\mathcal{H}_0\rightarrow\mathcal{H}$~\cite{Kruszynski}.
Then, using Theorem~2, if $\hat{C}^{\rm Y}$ and $\hat{L}^{\rm Y}$ are jointly measurable for either Alice or Bob, for each $c$ there exists a separable state which maximizes the expectation value of $\hat{L}_0$ (and hence, $\hat{L}$) over {\it all} quantum states making $\hat{C}$ and $\hat{L}$ unsuitable for UEW.

\subsection{Optimal Multipartite states}

Here we derive the optimal separable state for the simplest multipartite version of our protocol using the fixed three-outcome measurement scheme of Eq.~\eqref{POVMs}, and the test and constraint operators in Eqs.~\eqref{MultiTestOp} and~\eqref{MultiConstOp}, respectively.

First, the optimal separable states with respect to the partitioning $\mathbf{P}_k=(\mathcal{I}_1|\mathcal{I}_2|\cdots|\mathcal{I}_k)$must be of the form
\begin{equation}
|\psi_{\rm sep}\rangle = \bigotimes_{i=1}^k |\psi_i\rangle,
\end{equation}
where each $|\psi_i\rangle$ can be {\it entangled} for party $\mathcal{I}_i$.
It is clear that $\langle\psi_{\rm sep}|\hat{C}|\psi_{\rm sep}\rangle = c = c_1 c_2 \cdots c_k$ with $c_i=\langle\psi_i|\hat{C}_i|\psi_i\rangle$, and thus, for $c=0$, it is only required that one of the $c_i$'s equals zero.
Similarly, $\langle\psi_{\rm sep}|\hat{L}|\psi_{\rm sep}\rangle = g = g_1 g_2 \cdots g_k$ where $g_i=\langle\psi_i|\hat{L}_i|\psi_i\rangle$.
By extrapolating the bipartite case, it turns out that each vector $|\psi_i\rangle$ has the following form:
\begin{equation}
\begin{split}
&|\psi_u\rangle = f_u^{-\frac{1}{2}}\left[\bigotimes_{i=1}^{M_u}|\chi^{+}\rangle - (\frac{1-x}{2})^{\frac{M_u}{2}}\bigotimes_{i=1}^{M_u}|V\rangle\right],\text{~ for some party~}\mathcal{I}_u,\\
&|\psi_v\rangle = f_v^{-\frac{1}{2}}\bigotimes_{i=1}^{M_v}|\chi^{+}\rangle,\text{~ for all parties~}\mathcal{I}_v\neq \mathcal{I}_u,
\end{split}
\end{equation}
where $f_u$ and $f_v$ are appropriate normalizations.
The form of $|\psi_u\rangle$ guarantees that $c_u=0$ and thus $c=0$, while delivering the maximum value for $g_u$ within that partition.
The form of $|\psi_v\rangle$ guarantees that for each party $v$ the maximum $g_v$ will be obtained. To determine which partition must be chosen as party $\mathcal{I}_u$, we first calculate the following values.
\begin{equation}
g_u = (1-\frac{x}{2})^{M_u} - (\frac{1-x}{2})^{M_u}, \qquad g_v = (1-\frac{x}{2})^{M_v}.
\end{equation}
It can be easily verified that for $N_u=N_v$, $g_u<g_v$ and $g_u \rightarrow g_v $ as $N_u \rightarrow \infty$.
Hence, to obtain the maximum value $g$, we chose the party $u$ to be the party with the maximum number of agents (subsystems), i.e.,\ $\mathcal{I}_u=\mathcal{I}_k$ in the notation from the main text, where $M_1\leq M_2\leq\cdots\leq M_k$.
Consequently,
\begin{equation}\label{psiSep}
|\psi_{\rm sep}\rangle = f^{-\frac{1}{2}}\left[\bigotimes_{i=1}^{N-M_k}|\chi^{+ (i)}\rangle\right]\left[ \bigotimes_{i=1}^{M_k}|\chi^{+ (i)}\rangle - (\frac{1-x}{2})^{\frac{M_k}{2}}\bigotimes_{i=1}^{M_k}|V\rangle \right],
\end{equation}
with $f$ a normalization coefficient, and
\begin{equation}
g = g(x;N,M_k) = (1-\frac{x}{2})^{N-M_k}\left[ (1-\frac{x}{2})^{M_k} - (\frac{1-x}{2})^{M_k} \right].
\end{equation}
Note that, for $M_k=N$, i.e.,\ when no partitioning has been made, we find that the state $|\psi_{\rm sep}\rangle$ in Eq.~\eqref{psiSep} gives the highest expectation value for the test operator over the {\it full} state space. That is, the resulting state is genuinely entangled.

\end{widetext}
\end{document}